\documentclass[useAMS,usenatbib]{mn2e}
\usepackage[dvips]{graphicx}   % This is necessary to be able to include graphic

\usepackage[]{natbib,amsmath,amssymb}
\usepackage{latexsym,color,array}

\def\reff@jnl#1{{\rm#1\/}}
\def\aj{\reff@jnl{AJ}}                  % Astronomical Journal
\def\araa{\reff@jnl{ARA\&A}}            % Annual Review of Astron and Astrophys
\def\apj{\reff@jnl{ApJ}}                % Astrophysical Journal
\def\apjl{\reff@jnl{ApJ}}               % Astrophysical Journal, Letters
\def\apjs{\reff@jnl{ApJS}}              % Astrophysical Journal, Supplement
\def\apss{\reff@jnl{Ap\&SS}}            % Astrophysics and Space Science
\def\aap{\reff@jnl{A\&A}}               % Astronomy and Astrophysics
\def\aapr{\reff@jnl{A\&A~Rev.}}         % Astronomy and Astrophysics Reviews
\def\aaps{\reff@jnl{A\&AS}}             % Astronomy and Astrophysics, Supplement
\def\mnras{\reff@jnl{MNRAS}}            % Monthly Notices of the RAS
\def\prd{\reff@jnl{Phys.Rev.D}}         % Physical Review D
\def\prl{\reff@jnl{Phys.Rev.Lett}}      % Physical Review Letters
\def\pasp{\reff@jnl{PASP}}              % Publications of the ASP
\def\pasj{\reff@jnl{PASJ}}              % Publications of the ASJ
\def\nat{\reff@jnl{Nature}}             % Nature
\def\physrep{\reff@jnl{Phys. Rep.}}     % Physics Reports

\newcommand{\bea}{\begin{eqnarray}}
\newcommand{\eea}{\end{eqnarray}}
\newcommand{\be}{\begin{equation}}
\newcommand{\ee}{\end{equation}}
\newcommand{\rund}[1]{\left(#1\right)}
\newcommand{\vc}[1]{\mbox{\boldmath $#1$}}
\renewcommand{\d}{{\rm d}}
\newcommand{\dc}{\partial}

\newcommand{\abs}[1]{\left| #1 \right|}
\newcommand{\ts}{\thinspace}
\newcommand{\robs}{\langle r^2_{\rm obs}\rangle}
\newcommand{\rcal}{\langle r^2_{\rm cal}\rangle}
\newcommand{\rave}{\langle r_{\rm obs}\rangle}

\def\elabel#1{\label{eq:#1}}
\title[Halo ellipticity as a function of radius]
{Estimate of halo ellipticity as a function of radius with flexions}

\author%
[Er et al.]%
{Xinzhong Er$^{1}$\thanks{E-mail:xer@nao.cas.cn},
  Shude Mao$^{1,2}$,
  Dandan Xu$^{1}$,
  and
  Yixian Cao$^{1}$\\
  $^{1}$National Astronomical Observatory of China,
  Chinese Academy of Sciences, Beijing 100012,China\\
  $^{2}$Jodrell Bank Centre for Astrophysics, University of
  Manchester, Alan Turing Building, Manchester M13 9PL, UK}

\date{Accepted 2011 July 7; received 2011 June 11; in original form 2011 April 19}
\pubyear{2011}

\begin{document}

\maketitle

\begin{abstract}
  The cold dark matter theory predicts triaxial dark matter
  haloes. The radial distribution of halo ellipticity depends on
  baryonic processes and the nature of dark matter particles
  (collisionless or collisional). Here we show that we can use lensing
  flexion ratios to measure the halo ellipticity as a function of
  radius. We introduce a weight function and study the relationship
  between the first and second order statistics of flexion ratios,
  both of which can be used to reduce the bias in the estimate of
  ellipticity. We perform numerical tests for our method, and
  demonstrate that it can reduce the bias and determine
  the halo ellipticity as a function of radius. We also point out
  that the minimum mean flexion ratio can be used to trace the centres of
  galaxy clusters.
\end{abstract}
%%%%%%%%%%%%%%%%%%%%%%%%%%%%%%%%%%%%%%%%%%%%%%%%%%%%%%%%
\section{Introduction}
Structures of cluster- and galactic-sized dark matter haloes predicted
by N-body simulations of Cold Dark Matter (CDM) have several important
features, e.g. a nearly universal radial density profile for haloes over a wide
range of masses \citep{1996ApJ...462..563N,1997ApJ...490..493N},
and triaxial shapes
\citep{2002ApJ...574..538J,2004IAUS..220..421S}. These features are
related to the nature of dark matter particles, the formation processes
of galaxies and clusters, and their environments
\citep{2007ApJ...671.1135K, 2007ApJ...654...53M,
  2008ApJ...681.1076D,2009RAA.....9...41F,
  2010MNRAS.404.1137B,2010MNRAS.407..581R,2010arXiv1007.0612W,
2011arXiv1104.1566V}. For
example, major mergers of dark matter haloes may play an important role
in forming their shapes \citep{1993ApJ...418..544V}. Mismatches
between luminous galaxy shapes and total mass shapes
provide evidence for the existence of dark matter
\citep{2010A&A...524A..94S}. Numerical simulations under different
assumptions predict dark matter haloes with different shapes
\citep{2002sgdh.conf..109B, 2005ApJ...627..647B}. For instance,
collisionless CDM predicts prolate triaxial haloes
\citep{2006MNRAS.367.1781A}, while simulations with gas cooling
suggest more spherical shapes of dark matter haloes
\citep{2004ApJ...611L..73K}. Moreover, taking into account the
baryonic cooling effect, inner halo shapes are often found to be more
spherical than their outer counterparts
\citep{2004IAUS..220..421S,2004ApJ...611L..73K}.  In simulations
without baryonic cooling, however, the opposite trend has been seen,
i.e.  the outer parts of haloes appear to be rounder than inner parts
\citep{1991ApJ...378..496D}.

The shapes of cluster-sized haloes or galaxy-sized haloes can be studied with
different approaches, such as X-ray observations and the Sunyaev-Zeldovich
effect \citep{2000A&A...364..377R, 2004ApJ...601..599L,
  2004ApJ...617..847W, 2005ApJ...625..108D, 2006ApJ...645..170S}.
One can also use the spatial distribution of satellite galaxies
as a tracer for the shapes of their host dark matter haloes
\citep{2008MNRAS.385.1511W, 2010ApJ...718..762W}.

Gravitational lensing provides a powerful tool to study the mass
distributions of clusters of galaxies as well as galaxy haloes
\citep[see][for reviews on weak lensing]{2001PhR...340..291B,
  2003ARA&A..41..645R,2006glsw.book.....S, 2008PhR...462...67M}. This
is because gravitational lensing probes the distribution of matter
regardless of whether it is luminous or dark. Weak lensing techniques
have been used for cluster mass reconstructions
\citep[e.g.][]{2006ApJ...648L.109C,2008ApJ...687..959B} and for
determining halo ellipticities \citep{2004ApJ...606...67H,
  2009MNRAS.393.1235C, 2010ApJ...721..124D, 2010MNRAS.tmp..941H}. The
method of stacking galaxies to determine galaxy halo ellipticity was
proposed in \citet{2000ApJ...538L.113N} and has been used to determine
ellipticities of both cluster- and galactic-sized haloes
\citep{2009ApJ...695.1446E, 2006MNRAS.370.1008M}. For example, a mean
ellipticity of 0.46 has been found from a sample of 25 clusters using Subaru
data \citep{2010MNRAS.405.2215O}.

Flexion has been studied as the derivative of the surface mass density
and the shear, responds to small-scale variations in the projected
mass distribution
\citep{2002ApJ...564...65G,2005ApJ...619..741G,2006MNRAS.365..414B}.
Different techniques have been developed to measure flexion \citep{
  2006ApJ...645...17I, 2007ApJ...660..995O, 2007MNRAS.380..229M,
  2008A&A...485..363S, 2010arXiv1011.3041V,2011arXiv1101.4407F,
  2011arXiv1103.0551C}. Due to the complexity of deconvolving for the
point spread function, current lensing image measurements are not
accurate enough to fully study flexion. It has only been implemented
on limited samples of real data. \citet{2008ApJ...680....1O} and
\citet{2007ApJ...666...51L} have performed flexion measurements to
detect substructures in the cluster Abell 1689 using Subaru and HST
data. Though flexion noise is still poorly understood, flexion can
potentially contribute to cosmology in several aspects, such as
exploring dark matter haloes of galaxies and clusters, especially
substructures
\citep{2009MNRAS.395.1438L,2010MNRAS.409..389B,2010arXiv1008.3088E},
understanding the large scale structures through cosmic flexion
\citep{2010arXiv1003.5003M,2011arXiv1101.4769S}.
\citet{2011MNRAS.412.1023H} also proposed to reduce the distance
measurement errors of standard candles using lensing shear and flexion
maps. In \citet{2009MNRAS.400.1132H}, galaxy halo ellipticity has been
studied with flexion. Recently, \citet[][hereafter
  ES11]{2011A&A...528A..52E} developed a new approach using the ratio
of the tangential to radial flexion to estimate the halo ellipticity
as a whole. The approach is independent of lens strength, i.e. the
mass or the lens redshift. In particular, the flexion ratio is
independent of the halo-centric distance. In this paper, for the first
time, we use the flexion ratio to estimate how the halo ellipticity
varies with radius.

In reality, there are several complications. For example, the
background galaxy number density may not be sufficiently high, and
galaxies may not be circularly symmetrically distributed. Even worse
the intrinsic and other noises will cause an overestimate for the mass
ellipticity. In this paper, we introduce a weight function and a
method that combines the first and second order statistics of flexion
ratios to reduce the bias (see Appendix A), we test our method
with numerical simulations (Sect. 3). Our results are discussed in
Sect. 4.
The cosmology that we adopt throughout this paper is a
$\Lambda$CDM model with $\Omega_{\Lambda}=0.75$, $\Omega_{m}=0.25$,
$\sigma_8=0.9$ and a Hubble constant $H_0=73\, {\rm km}\,{\rm s}^{-1}{\rm Mpc}^{-1}$.

\section{\label{Sc:2}Basic formalism}

The full formalism described here can be found in \citet{2006MNRAS.365..414B,
2008A&A...485..363S}.
Weak lensing shear and flexion are conveniently described
using a complex formalism. We adopt the thin lens approximation,
assuming that the lensing mass distribution is projected onto a single lensing
plane. The dimensionless projected mass density (convergence) can be written as
$\kappa(\vc \theta)= \Sigma(\vc\theta)/ \Sigma_{\rm cr}$,
where $\vc\theta$ is the position coordinate, $\Sigma (\vc\theta)$ is the
projected mass density and $\Sigma_{\rm cr}$ is the critical density, given by
\be
\Sigma_{\rm cr}= {c^2 \over 4\pi G} {D_{\rm s}(z_s)\over
D_{\rm d} D_{\rm ds}},
\ee
for a fiducial source located at redshift $z_s$. Here
$D_{\rm s}$, $D_{\rm d}$ and $D_{\rm ds}$ are the angular
diameter distance between the observer and the source, the observer and the
lens and between the lens and the source, respectively.

The first-order image distortion is measured by shear
$\gamma$, which transforms a circular source into an elliptical one. The
second order effect, which is called flexion, is described by two
parameters.  The spin-1 flexion is the complex derivative of $\kappa$
\be
{\cal F}= \nabla_{\rm c} \kappa= {\dc \kappa \over \dc\theta_1}
+ {\rm i}{\dc\kappa \over \dc\theta_2},
\ee
and the spin-3 flexion is the complex derivative of $\gamma$
\be
{\cal G}= \nabla_{\rm c} \gamma.
\ee
\subsection{Flexion ratio}

The full definition of the flexion ratio can be found in ES11, here we
only briefly describe the essential quantities. The flexion ratio is
given by
\be
r={{\cal F}_T\over {\cal F}_R},
\elabel{fratio}
\ee
where ${\cal F}_T$ and ${\cal F}_R$ are the tangential and radial
components of the spin-1 flexion ${\cal F}$, respectively. ES11 has
shown that for an elliptical isodensity contour of the mass
distribution, the flexion ratio provides an elegant estimator for the
halo mass ellipticity $\epsilon$ and the orientation angle $\phi_0$
\be
r(\phi)={|2 \epsilon \sin 2(\phi-\phi_0)| \over 1-2\epsilon\cos2(\phi-\phi_0)
  +\epsilon^2},
\elabel{ratio-phi}
\ee
where $\phi$ is the polar angle,
$\epsilon=(a-b)/(a+b)$ and $a$, $b$ stands for the major and minor axes
of the projected halo ellipse respectively.
The expectation value of the flexion ratio is thus directly related to
the halo ellipticity
\bea
\langle r\rangle &=& {1\over 2\pi} \int_0^{2\pi} \d \phi
\abs{\frac{2\epsilon\sin 2(\phi-\phi_0)}{1- 2\epsilon \cos 2(\phi-\phi_0)
    +\epsilon^2}} \nonumber\\
&=&{2\over \pi}\, {\rm ln}{1+ \epsilon  \over 1-\epsilon}\;.
\elabel{rmean}
\eea
The second order moment of $r$ depends only on $\epsilon$
\bea
\langle r^2\rangle &=& {1\over 2\pi}\int_0^{2\pi}\d \phi
\rund{2\epsilon\sin2(\phi-\phi_0) \over
1-2\epsilon\cos2(\phi-\phi_0)+\epsilon^2 }^2\\
&=& {2\epsilon^2 \over 1-\epsilon^2}.
\elabel{r2mean}
\eea
We then have two estimators $\hat\epsilon$ and $\hat\epsilon'$ for
the halo ellipticity using $\langle r\rangle$ and $\langle r^2\rangle$
\bea
\hat \epsilon= \dfrac{{\rm e}^{\pi \langle r\rangle /2}-1}
{{\rm e}^{\pi \langle r\rangle/2}+1};
%\;\; {b\over a}={1\over {\rm e}^{\pi \langle r \rangle/2}};
\elabel{estimator1}\\
\hat \epsilon'=\sqrt{\langle r^2\rangle \over 2+\langle r^2\rangle}.
\elabel{estimator2}
\eea
Due to the intrinsic and other noises, both estimators are biased.
The bias behaviors in $\langle r\rangle$ and $\langle r^2\rangle$ are
different. By comparing $\langle r\rangle$ and $\langle r^2\rangle$,
we can obtain an approximation for the bias $b_r$
(Eq.\ref{eq:ratiobias}), which we use to subtract from the observed
$\robs$ by
\be
\langle r_{\rm clean}^2\rangle =\robs -2\langle r\rangle b_r.
\ee
Notice the bias $b_r$ is not zero since it is not the same as the average
noise, which is zero.
Using $\langle r_{\rm clean}^2\rangle$ in Eq.(\ref{eq:estimator2}), a
``clean'' estimator is obtained, and will be employed in the
following section.  A detailed description of bias reduction is given
in Appendix A2.
\section{Mass ellipticity as a function of radius}

In this section we present how we estimate mass ellipticities at
different radii using numerical simulations. For an elliptical
isodensity contour mass distribution, the flexion ratio $r$ is independent
of the angular distance to the centre of the lens $\theta$.
%It can be easily seen that for a power-law radial profile
%$\kappa\sim\theta^{-\beta}$, both radial and tangential flexion will
%vary with the same profile $\sim\theta^{-(\beta+1)}$.
Therefore the ellipticity as a function of $\theta$ can be
estimated. Several annular bins could be easily applied to the
data. However, at small radii, the number of background galaxies in
each bin is low, since it is proportional to $\theta$ for a fixed bin
width. On the other hand, the flexion signal drops rapidly at large
$\theta$, $\propto \theta^{-2}$ in an isothermal halo, and even faster
in an NFW halo, approximately $\propto \theta^{-3}$
($\theta>r_s$, where $r_s$ is the scaling radius).
One might expect a better signal-to-noise (S/N) ratio
at smaller radii. In reality, both are biased for different
reasons. At small radii we obtain an underestimate due to the lack of
background galaxies, and at large radii the estimate is strongly affected
by noise, which results in overestimates.

Shapes of dark matter haloes have been studied by many authors using
numerical simulations
\citep{2002ApJ...574..538J,2002sgdh.conf..109B,2004IAUS..220..421S,
  2006MNRAS.367.1781A}. In these studies, there are different
ways of modeling haloes as ellipsoids. One popular method uses
some form of eigenvectors of the inertia tensor.
%The eigenvectors correspond to the direction of the major axes, and
%the eigenvalues correspond to the length of the semi-major axes $c\leq
%b\leq a$, from which the axial ratios are defined.
Another method fits the isodensity surfaces of the halo as ellipsoids
\citep{2002ApJ...574..538J}. In this paper, we employ the latter method.
We fit the $n$th isodensity contour using the $n$th
surface mass density $\kappa_n$ and estimate the major axis $a_n$ and
minor axis $b_n$. We then calculate the ellipticity using
$(a_n-b_n)/(a_n+b_n)$ at the ``effective'' radius $\sqrt{a_n b_n}$. The
ellipticity derived using the flexion ratio approach will be compared
with the ellipticity calculated in this way.
\subsection{\label{sc:nfw}A halo modelled by an NFW profile}
\begin{figure}
  \includegraphics[angle=0,scale=1.3]{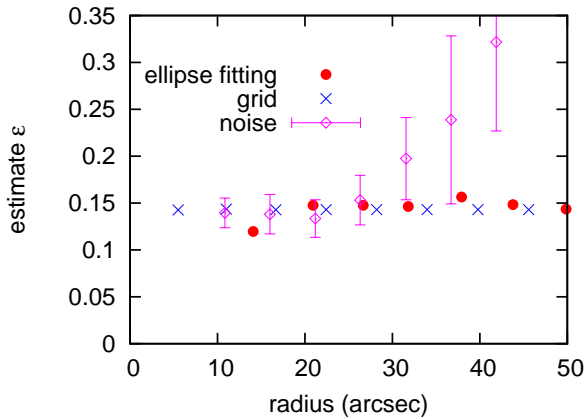}
  \caption{Comparison of different halo ellipticity estimators using an
  NFW halo. The filled circles are estimated directly from
    fitting with surface mass ellipse, the crosses are estimated from
    the flexion ratio method (Sec.\ts3.2) with noise-free grid-based
    data. The diamonds are estimated using the clean method but with
    noisy data, and they are mean result over 50 realisations, the
    errors represent the standard deviations.}
  \label{fig:nfw}
\end{figure}

We first generate an a halo modelled by the NFW (Navarro et al. 1996, 1997)
profile with constant ellipticity. The
dimensionless surface mass density is written as
\citep{1996A&A...313..697B, 2006MNRAS.365..414B}
\be
\kappa(x)=2\kappa_s{f(x)\over x^2-1},
\elabel{kappa}
\ee
where $x$ is the radius in units of the scaling radius $r_s$, such that
$x=\theta/r_s$ and $\theta=\sqrt{\theta_1^2/(1+\epsilon)^2 +
\theta_2^2/ (1-\epsilon)^2}$. The ellipticity we choose is
$\epsilon=0.15$. We define
$\kappa_s=\rho_{\rm crit} r_s \Delta_c/\Sigma_{\rm cr}$, where
$\Delta_c$ is the dimensionless characteristic density
(see Appendix in Navarro et al. 1997).
The lens halo is at redshift $z_{\rm d}=0.6$ and sources at
$z_s=1.48$.
The halo has mass $M_{200}=1.8\times10^{14} h^{-1}M_{\odot}$ and
concentration parameter $c=7.2$ (the corresponding scaling radius
$r_s= 27$ arcsec).
The function $f(x)$ is given by
\be
f(x)=
\begin{cases}
  1 - \dfrac{2}{\sqrt{x^2 - 1}}{\rm arctan}\sqrt{\dfrac{x-1}{x+1}} \;\; (x>1);\\
  \\0 \quad\quad\quad\quad\quad\quad\quad\quad\quad\quad\quad \;\; (x=1); \\ \\
  1 - \dfrac{2}{\sqrt{1-x^2}}{\rm arctanh}\sqrt{\dfrac{1-x}{1+x}} \;\; (x<1). \\
\end{cases}
\ee
The flexion data are generated by
\be
{\cal F}(x)=-{2{\cal F}_s\over (x^2-1)^2}[2xf(x)-h(x)]
(\cos\phi+{\rm i}\sin\phi),
\ee
where ${\cal F}_s = \kappa_sD_{\rm d}/r_s$. The polar angle
can be determined from $\tan\phi=\dfrac{\theta_2 (1+\epsilon)^2}
{(1-\epsilon)^2 \theta_1}$, and
\be
h(x)=
\begin{cases}
1-\dfrac{2x}{\sqrt{x^2-1}}{\rm arctan}\sqrt{\dfrac{x-1}{x+1}}-1/x \;\;(x>1);\\
\\-1 \quad\quad\quad\quad\quad\quad\quad\quad\quad\quad\quad \;\; (x=1); \\ \\
1-\dfrac{2x}{\sqrt{1-x^2}}{\rm arctanh}\sqrt{\dfrac{1-x}{x+1}}-1/x \;\;(x<1).\\
\end{cases}
\ee
Two sets of mock data are generated. First we generate flexion data on
$100\times100$ grids in a field of $2\times2$ arcmin$^2$. In the other
data set we have 1000 flexion data on random locations in the same
field. We add the intrinsic Gaussian flexion noises $0.03$ arcsec$^{-1}$ for the
$\theta_1$ and $\theta_2$ components respectively (more details are
given in Appendix B). The data are discarded if they are located too
close to the halo centre, $\theta < \theta_{\rm min}=6$ arcsec or too
far from the halo centre, $\theta>\theta_{\rm max}=55$ arcsec. Since
a very large flexion cannot be measured \citep{2008A&A...485..363S},
data points with $|{\cal F}|>0.5$ are also discarded. Moreover, due to the
upper bound of flexion ratio (ES11), those with $r>4.5$ are discarded
as well (in all the numerical tests below, similar filters but
different $\theta_{\rm max}$, $\theta_{\rm min}$ are employed). The
ellipticity $\epsilon$ is also estimated using the surface mass
density $\kappa$ as a comparison. We apply 8 bins in performing the
flexion ratio method. In each annulus, an estimate for $\epsilon$ is
obtained, as well as the radius, taken as the mean radius over all the
data in the annulus. The result is shown in Fig.\ts\ref{fig:nfw}. For
data with noise, 50 realisations are used. The mean values and standard
deviations of our estimates over 50 realisations are shown by the
diamonds and error bars. One can see that the ellipticity estimates
using $\kappa$ (filled circles) and grid noise-free data (crosses)
closely trace our input values. The results using data with noise have a
mean error $<10\%$ at small radii ($<30$ arcsec).  However for the
large radii, the results are dominated by the intrinsic noise, and thus are
significantly overestimated as expected.

\subsection{A galactic-sized halo}
We take a galaxy halo from the Aquarius project
\citep{2008MNRAS.391.1685S}, which is a suite of $N$-body simulations
of galaxy-sized CDM haloes. The halo Aq-F at resolution level 2 is
selected for our test, with a softening length of $\sim 0.05 h^{-1}$
kpc, mass resolution of $5\times 10^{3} h^{-1} M_{\odot}$. The main
halo contains about $1.4 \times 10^8$ particles -- a total mass of
$7.2 \times 10^{11} h^{-1} M_{\odot}$ within its virial radius of $153
h^{-1}$ kpc. We place the halo at a lens redshift $z_d=0.6$. The halo
is then projected along 20 different directions to generate 20
different sets of mock data. The background sources are randomly
located in a ``source plane'' at redshift $z_s=1.0$. The $\kappa$ and
flexion maps are generated on $100\times100$ grids in the lens
plane. The flexion for each image is linearly interpolated using
values of the four nearest grid points. Furthermore, we add
Gaussian intrinsic flexion noise ($0.03$ arcsec$^{-1}$) to the data.
%rs=r200/c, c=16.2

\begin{figure}
  \centering \includegraphics[angle=0,scale=1.3]{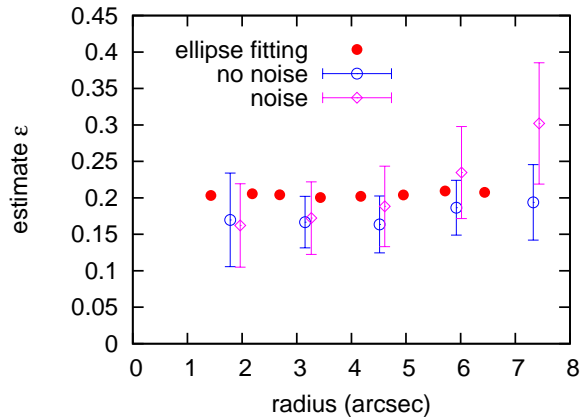}
  \caption{Similar to Fig.\ts\ref{fig:nfw}, but for results from
    simulated galaxy-galaxy lensing. The circles are estimated using
    the flexion ratio method (Eq.\ref{eq:rmean}) using noise free
    data, and the diamonds are estimated using noisy data. The errors
    represent the standard deviations over the 50 realisations.}
  \label{fig:stack}
\end{figure}

To test the ellipticity as a function of radius, we perform our method
on stacked galaxy-galaxy lensing since the number density of
background galaxies is not sufficient to study an individual
galaxy. In each mock data projection, we use 10 flexion at random
locations in an area of $20\times20$ arcsec$^2$ (which is equal to 90
background galaxies/arcmin$^2$). The data from 20 different
projections are aligned and stacked with repect to the centre of the
lens halo. 50 realisations are generated to calculate the mean
value. Similar filters as in the last section are employed here,
except that we discard data closer than 1 arcsec o the halo centre, or
further than 8 arcsec away. The ellipticity using the surface mass
density is calculated, and the average ellipticity over 20 projections
is shown by the filled circles in Fig.\ts\ref{fig:stack}. The open
(blue) circles are the estimates from the flexion ratio method using
data without intrinsic noise and the diamonds (purple) are those using
noisy data. They are both averages over 50 realisations. The error
bars represent the standard deviations in each bin. The error bars
using noise free data in Fig.\ts\ref{fig:stack} are caused by the
variances of halo ellipticities projected from different directions,
while the error bars using noisy data are affected by the estimate
error. We can see that the error bars increase with radius, as
intrinsic noise becomes more important.
\subsection{A cluster-sized halo}
In addition, we test our approach on an $N$-body simulated galaxy
cluster halo. For galaxy clusters, we have more background galaxies in
``observations''. Therefore the flexion number density is high enough
to resolve an individual cluster. The test cluster halo PH-A-2 is
taken from the PHOENIX simulations (Gao et al. 2011 in prep). At the
second resolution level, the simulation has a softening length of
$0.32 h^{-1}$ kpc, mass resolution $5 \times 10^{6}
M_{\odot}h^{-1}$. The main halo contains $1.5\times 10^8$ particles,
with a total mass of $6.6\times10^{14}M_{\odot}h^{-1}$ within a virial
radius of $1.4h^{-1}$ Mpc. Two simulated cluster haloes are used to
generate mock data: one has only the main halo, while the other one in
addition includes massive subhaloes that are likely to have noticeable
signatures. The number of subhaloes is about $5\%$ of the total
subhaloes with mass $M>10^{9}M_\odot$ that are identified from the
simulations. The halo is located at $z_d=0.5$, while the background
sources are chosen to be at $z_s=1.0$. We project the same halo along
3 different directions to generate mock data.

\begin{figure}
  \centerline{\scalebox{1.1}
  {\includegraphics[width=7cm,height=7cm]{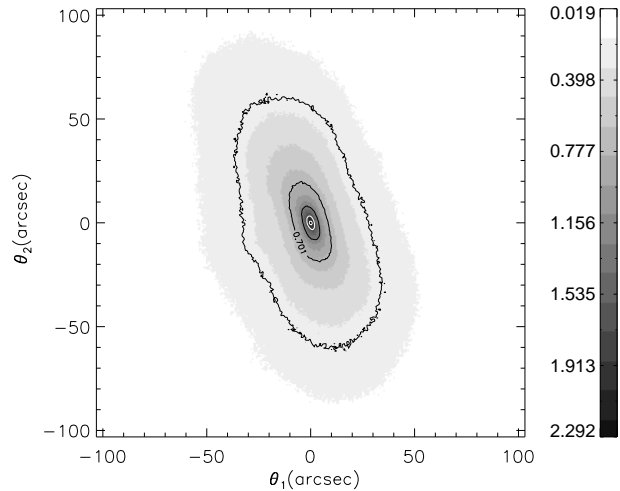}}}
  \caption{The surface density $\kappa$ map of the simulated cluster taken from
    the PHOENIX simulations. This halo is used for generating our flexion
    data. }
  \label{fig:kappa}
\end{figure}

\begin{figure*}
  \centering
  \includegraphics[angle=0,scale=1.15]{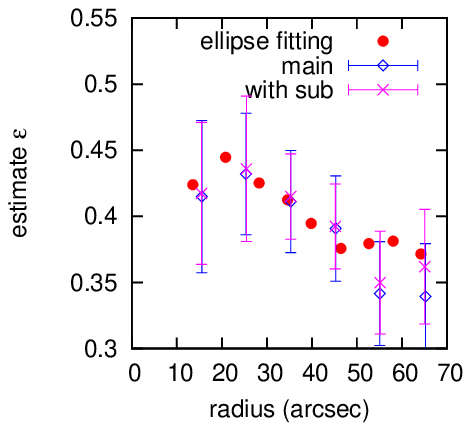}
  \includegraphics[angle=0,scale=1.15]{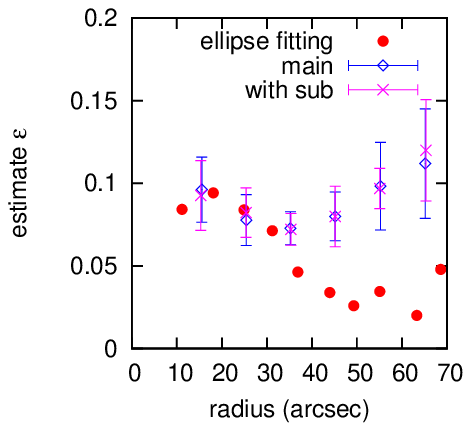}
  \includegraphics[angle=0,scale=1.15]{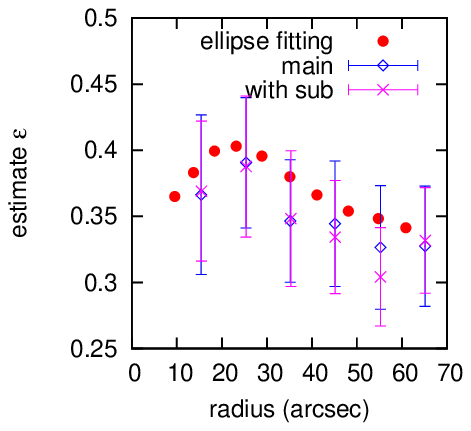}
  \caption{ Same as Fig.\ts\ref{fig:nfw} for the mock galaxy cluster
    from the PHOENIX simulations projected along the x-axis
    (left),y-axis (middle) and z-axis (right). The diamonds (crosses)
    are the mean result over 50 realisations using halo without (with)
    subhaloes.}
  \label{fig:cluster}
\end{figure*}

In Fig.\ts\ref{fig:kappa}, the surface mass density $\kappa$ map from
one projection (for the right panel in Fig.\ts\ref{fig:cluster}) is
shown. One can see that there are some small scale fluctuations due to
substructure. Moreover, the shape of the halo is not perfectly
elliptical. It is easy to see that there is a bump on the left side of
the halo. Both of these will cause some difficulties in estimating the
halo ellipticity since flexion is sensitive to small scale
variations. In each direction, we have 2 sets of data generated from
the cluster haloes with and without subhaloes.  We have 50
realisations with each having 800 background flexion data points on an
area of $200\times200$ arcsec$^2$. Such a configuration 
is available via the current surveys, e.g. the
COSMOS survey has a background density of 76 galaxies/arcmin$^2$
\citep{2010A&A...516A..63S}. Similar filters as in previous sections
with $\theta_{\rm min}=10$ arcsec and $\theta_{\rm max}=70$ arcsec are
applied here. About 300 data points are obtained on average in each
realisation. We make 6 bins in each realisation, and estimate the mean
ellipticity and its standard deviation in each bin over the 50
realisations. The results are shown in Fig.\ts\ref{fig:cluster}. The
filled circles are directly estimated from the surface mass density
without substructure. The diamonds are the results from the data with
the main halo, and the crosses are the results from the data with the
main halo and the substructures. We can see that our estimates are
slightly biased. The large error in the middle panel is caused by the
intrinsic noise, which has very small ellipticity.
In an additional test, we perform our estimates with a background galaxy 
number density of $45$ galaxies/arcmin$^2$ and obtain a similar result
but with slightly larger errorbars ($\sim 10\%$).

\begin{figure}
  \centering
  \includegraphics[angle=0,scale=1.4]{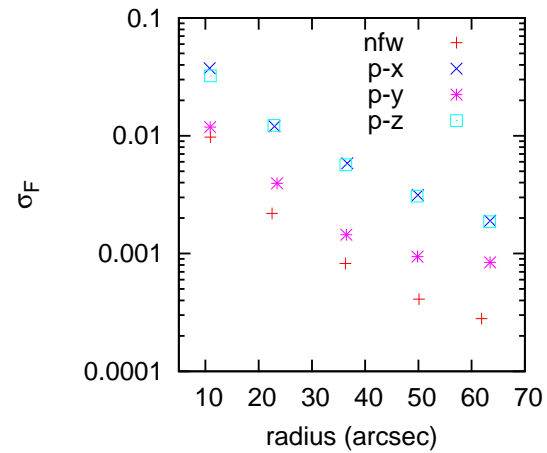}
  \caption{The standard deviation of the tangential flexion in annuli
    for the elliptical NFW halo (pluses) and the PHOENIX halo in 3
    different projections, in units of arcsec$^{-1}$.}
  \label{fig:sigma}
\end{figure}

There are however some other reasons that can cause bias in our
estimate. First of all, we need to point out that the estimate
directly from mass is not the ``true value'' of the ellipticity. When
we fit the iso-density contours using $\kappa$, we notice that the
orientations of different annuli are not aligned. Moreover, the annuli
are not concentric, especially for clusters. Although our estimate
using the flexion ratio does not depend on the radius or orientations,
the binning of the data will be affected by the rapid variation of
ellipticity and orientation. Secondly, the intrinsic noise and
substructures will usually cause an overestimate (ES11). As pointed
out by \citet{2010MNRAS.409..389B}, the galaxy-galaxy lensing flexion
varies due to substructures. As an additional test, we calculate the
standard deviations of tangential flexion in each bin for our NFW halo
and the PHOENIX halo with substructures. In Fig.\ts\ref{fig:sigma},
the values indicated by pluses are significantly smaller than the
others since the NFW halo has regular shape and contains no
substructures. The others from the PHOENIX halo have higher deviation
as we expect. We can also see that the ellipticity also affects the
deviations.

\begin{figure}
  \centering
  \includegraphics[angle=0,scale=1.4]{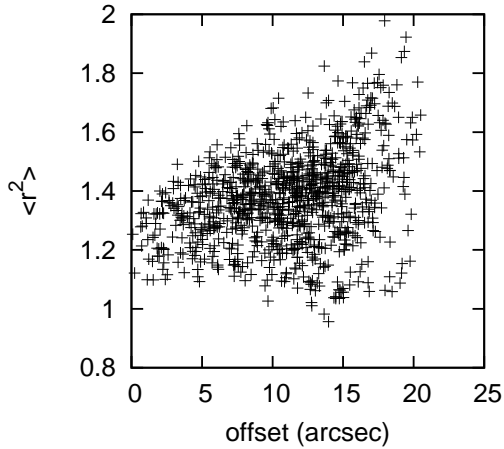}
  \caption{
    $\robs$ distribution due to the centroid offset for 10 realisations.
    For each realisation, 100 random offsets are used.
  }
  \label{fig:offset}
\end{figure}

The centres of galaxy clusters are often not well determined from
observations, e.g. it is often assumed that the mass centroid
coincides with the brightest cluster galaxy, which may not be correct
in all cases. Also mass and light peaks are often offset in the
weak lensing shear method \citep{2011arXiv1103.4607D}.  On the other
hand, in ES11 it has been found that a centroid offset to the true
mass centre of the halo causes a slight overestimate of the halo
ellipticity. The reason is that the flexion ratio is defined with
respect to the true centre of halo mass, a centroid offset will
introduce extra asymmetry in the estimate. Therefore, one can use the
flexion ratio as a probe for the cluster centre, since $\langle
r\rangle$, $\robs$ or an estimate of ellipticity, in principle, will
reach its minimum when the flexion ratio is calculated with respect to
the true mass centre. However the intrinsic and other kind of noises
can cause some complications. In order to see this, we perform another
test for the centroid offset problem. We generate 10 realisations of
mock data. For each realisation, we use 100 random mock centres within
20 arcsec to the true mass centre. The flexion ratio and $\robs$ are
calculated with respect to the mock centre, and the result is shown in
Fig.\ts\ref{fig:offset}. One can see that although $\robs$ does not
reach minimum at the true centre of the cluster, the variance of
$\robs$ monotonically increases with the offset.  This is also
because, as we mentioned in the last paragraph, the mass distribution
of galaxy clusters is complicated, e.g. the mass centre may vary with
the radius. Nevertheless, the minimum flexion ratio statistics
provides a complementary approach for cluster centre determination.

\section{Conclusions and Discussions}

In this paper, we have studied the flexion ratio method for estimating
shapes of elliptical dark matter haloes. We performed numerical
simulations to study our ellipticity estimators and obtained the halo
ellipticity as a function of radius. In particular, for galactic-sized
haloes, we are able to measure the mean variation of ellipticity with
radius by stacking several galaxy samples. For cluster haloes, we can
estimate the ellipticity as a function of radius for an individual
galaxy cluster. We also point out that the minimum $\robs$ provides
another approach for determining the centre of galaxy clusters.
Moreover, we introduced a weight function (in Appendix) to suppress
the bias due to the asymmetric distribution of background
galaxies. The variance of our estimate becomes significantly smaller
after employing the weight function. We derived the second order
statistics of the flexion ratio $\langle r^2\rangle$ and its
analytical relation to $\langle r\rangle$. By combining $\langle
r\rangle$ and $\langle r^2\rangle$, in principle the bias due to
intrinsic noise can be subtracted. We provided a method to reduce the
bias based on a linear approximation (For more details, see
Eq. \ref{eq:delta}).

We however notice that there are some difficulties in estimating the halo
ellipticity.
\begin{itemize}
\item
  First of all, the noise model that we employ in this paper
  includes only intrinsic noise and numerical noise in the
  simulations. In reality there are other sources of noise,
  e.g. noises due to the point spread function and
  pixelisation. Moreover, the noise of flexion measurement appears
  different using different methods \citep{2007ApJ...660.1003G}. Thus
  the noise in practice may be non-Gaussian. Therefore, the results
  presented here are based on somewhat idealised observations.
\item
The difficulty in reducing bias for
small ellipticity haloes is mainly due to the lack of knowledge for the
flexion noise behaviours. Moreover, the linear approximation that we
employed is based on the assumption that the bias is independent of
the halo ellipticity and is smaller than the flexion ratio signal. In
reality, neither of these assumptions may be valid. In particular, for
small elliptical haloes, our measurement will be dominated by
intrinsic and other noises.
\item
The optimal binning to the data is
also difficult since we do not know how the ellipticity changes with
radius. A coarse binning can only estimate the ``mean''
ellipticity. Moreover, as we show in the figures, flexion data on
different radii have different signal-to-noise ratio (S/N), thus better data
at small radii may be polluted by the low S/N data at large radii. On
the other hand, if many bins are used, not enough galaxies will be contained
within each bin.
%Perhaps an intermediate radius will be the promising region where we
%can obtain mass ellipticity estimate. It is however difficult to
%determine the region due to our lack of knowledge about the magnitude
%and behavior of intrinsic flexion noise. Moreover, there is no unique
%way to define the ``radius'' for ellipsoids. The method that we employed
%$r=\sqrt{a b}$ is not consistent with the method that we
%perform on the binned flexion data.
\item
The degeneracy of substructures and halo ellipticity: substructures
behave similarly to intrinsic flexion in our estimate, but at
different magnitudes. Substructures are therefore another kind of
``noise''. On the other hand, flexion can be used to measure
substructures in galaxies or clusters (e.g. their mass function and
spatial distribution). Fig.\ts\ref{fig:sigma} is a rough estimate by
comparing the $\sigma_F$ magnitude without any knowledge of
substructures or ellipticity. We can see that the combination of
different approaches may constrain both shapes and substructures in
dark matter haloes. Again good knowledge of flexion noise will be
necessary. Also, a realistic model of the substructure mass function, and
the spatial distribution of substructure will be helpful.
\end{itemize}

A high number density of background galaxies is desirable.  Current
surveys such as COSMOS may have sufficiently high galaxy number
densities to enable our method to constrain mass ellipticities on a
few radial bins for an individual galaxy cluster. This is again based
on a simple noise model. In reality, the noise of high magnitude
images is larger. The number density of high quality images using
current observations may not as high as we used. For galaxy-sized
haloes, we can perform stacking galaxy-galaxy lensing. Notice that
different from the shear method, we do not need to align the major
axis of the foreground galaxies.  We do, however, assume that we can
select dark matter haloes with similar shapes by carefully choosing
the foreground galaxies according to their luminosity, colour and
shape.
%That will require similar morphology of lensing galaxies, and
%rescaling lensing galaxies if they have different sizes or are
%located at different redshifts.
The different morphologies will cause uncertainty in our estimate. On
the other hand, the number density of galaxy-galaxy pairs will be very
large in current and future surveys, i.e. the number of stacked
galaxies in reality can be much higher than what we used in our
test. Fig.\ref{fig:stack} indicates we can measure $\epsilon$ to
standard deviation $\sigma_{\epsilon}\approx 0.05$ by stacking 20
foreground galaxies for $\epsilon\approx 0.2$. With a much larger
number of galaxies available in future observations, the accuracy of
this measurement can be improved. The improvement will be Poissonian
if there are no other sources of systematic errors. An accurate
measurement of how the ellipticity varies with radius will provide a
strong test of galaxy formation models.

\section*{Acknowledgments}
We would like to thank Richard Long, Yougang Wang, Lin Yan and an anonymous
referee for useful comments on the manuscript. We also thank Liang Gao
for help on the PHOENIX simulation data.

\appendix
\section[]{Bias estimate and reduction}
\label{sec:appendix}
In this appendix, we will introduce two approaches that will reduce the
estimated error.
\subsection{Weight function}
One of the biases comes from the asymmetric distribution of
background galaxies. To account for such effects, we introduce a weight
function
\be
w_i={1\over A N_i},
\elabel{weight}
\ee
where $A$ is a normalisation, and $N_i$ is the number of background
galaxies in a polar angle range $\Delta{\phi}$ centred at the $i$th data.
Therefore, the data is down-weighted where the number density is high. One
can see that if the galaxy images are circularly symmetrically distributed
with respect to the foreground lens galaxy/cluster, the weight
function will be uniform $w_i=1/N$, where $N$ is the total number of
background galaxies. The weight function is employed when calculating
the expectation flexion ratio from data
\bea
\bar r &=&\sum_{i=1}^{N} r_i w_i;
\elabel{weightr}\\
\bar {r^2}&=&\sum_{i=1}^{N}r^2_i w_i.
\elabel{weightr2}
\eea
\begin{figure}
  \centering
  \includegraphics[angle=0,scale=1.5]{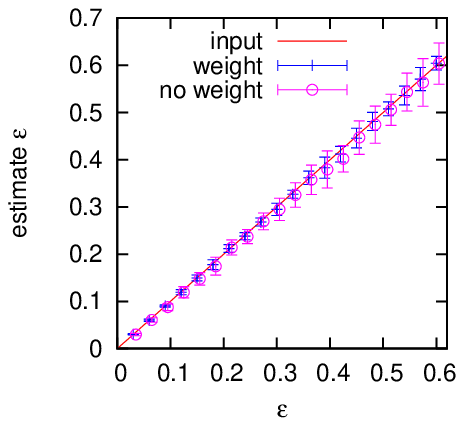}
  \includegraphics[angle=0,scale=1.5]{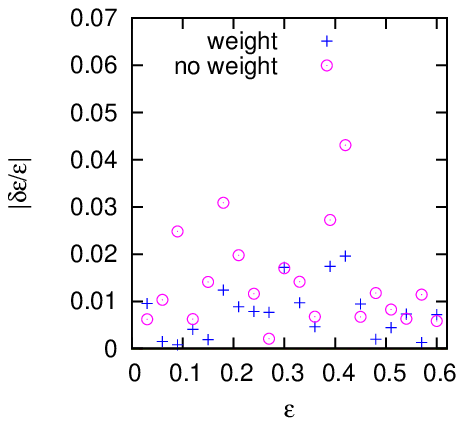}
  \caption{Comparison of the ellipticity estimators with/without the weight
    function. In the top panel, the solid line is the input
    ellipticity, the pluses are the results using the weight function
    (Eq.\ref{eq:weight}), and the circles are the results without the
    weight function. The error bar is calculated by the standard
    deviation. The bottom panel represents the relative errors.
  }
  \label{fig:weight}
\end{figure}
\subsection{Bias reduction}
Due to the difficulty of measurement and our incomplete understanding of
the point spread function (PSF), our knowledge about the intrinsic
flexion is not sufficient to give an accurate noise model. Since
flexion has a dimension of inverse length, the intrinsic flexion
depends on the image size, and therefore the depth of
survey. Moreover, \citet{2007ApJ...660.1003G} found that different
methods of measurement may lead to different noise behaviors.

\begin{figure}
  \centering
  \includegraphics[angle=0,scale=1.5]{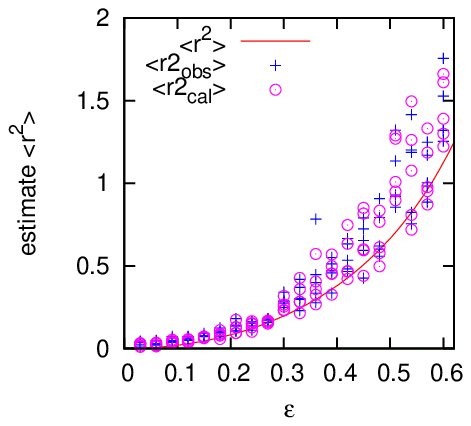}
  \caption{Comparison of $\robs$ and $\rcal$ with $\epsilon$: The
    solid line is for theoretical value of $\langle r^2\rangle$
    (Eq.\ref{eq:r2mean}), the pluses are obtained directly from
    data (Eq.\ref{eq:weightr2}), and the circles are calculated
    using Eq.(\ref{eq:r2ofr}), with $\langle r\rangle$ obtained
    from data (Eq.\ref{eq:weightr}).}
  \label{fig:r2dif}
\end{figure}

In ES11, it has been found that estimates using flexion ratios will
be larger than the true ones due to the intrinsic noise. Here we adopt
a simple model for the bias in the flexion ratio $r$, by assuming that
the bias is independent of the halo ellipticity and lens strength.
The observed mean flexion ratio can be written as
\be
\langle r_{\rm obs} \rangle = \langle r \rangle +b_r,
\elabel{noiser}
\ee
where $b_r$ stands for the bias in the $\langle r\rangle$
estimate.  Notice that $b_r$ is not zero because the noise of the
flexion ratio is not Gaussian (even if we assume that the noise of two
flexion components is Gaussian, the ratio of two Gaussians
no longer follows a Gaussian distribution). $\langle
r^2\rangle$ can be obtained through two ways. First, we can
directly calculate $\robs$ from data using Eq.(\ref{eq:weightr2}).
Secondly, according to Eqs.(\ref{eq:rmean}) and
(\ref{eq:r2mean}), we have
\be
\langle r^2_{\rm cal}\rangle={({\rm e}^{\pi \langle r\rangle /2}-1)^2 \over 2
  {\rm e}^{\pi \langle r\rangle /2}}.
\elabel{r2ofr}
\ee
We can thus calculate $\rcal$ using Eq.(\ref{eq:r2ofr}) from $\langle
r_{\rm obs}\rangle$. The bias strength will be different in these
two methods (Fig.\ts\ref{fig:r2dif}). One can see that both estimates
give larger values than the theoretical prediction (solid line), and they
are different from each other. We make a first order expansion
\bea
\robs &=& \langle r^2\rangle
+2\langle r\rangle b_r +O(b_r^2);\\
\rcal &=& \langle r^2\rangle +{\dc \langle r^2\rangle \over
  \dc \langle r\rangle} b_r + O(b_r^2)\\
&\approx& \langle r^2\rangle + {\pi \over 4}
\rund{{\rm e}^{\pi \langle r\rangle /2} - {\rm e}^{-\pi \langle r\rangle /2}} b_r.
\elabel{noiser2}
\eea
The difference between $\robs$ and $\rcal$ provides an estimate of
the bias $b_r$,
\bea
\Delta &=&\robs-\rcal \\
&\approx& \rund{2\langle r\rangle- {\pi \over 4}
\rund{{\rm e}^{\pi \langle r\rangle /2} - {\rm e}^{-\pi \langle r\rangle /2}}}
b_r + O(b_r^2).
\elabel{delta}
\eea
We can subtract $b_r$ from the estimate, and obtain a relatively clean
result. However, this only works for large $\epsilon$. For small
$\epsilon$ {\bf($\lesssim 0.1$)}, our approximation is no
longer valid, since our measurements are dominated by noise.  In the case of
small $\epsilon$, the $b_r^2$ term is larger than the
other terms in Eq.(\ref{eq:noiser2}), and thus cannot be
neglected. Due to incomplete knowledge of the flexion noise, it is
difficult to estimate when we should take the $b_r^2$ term into
consideration.
%We only make a trade off for the noise reduction. In principle, our
%estimated noise $\delta_r'$ must be smaller than $\rave$,
%i.e. $\langle r\rangle$ is always positive. Once we obtain a
%$\delta_r'$ larger than $\rave$, we will use $2\Delta$ as an
%empirical approximation for the $\delta_r^2$ term, subtract it from
%$\robs$, and use Eq.(\ref{eq:r2mean}) to estimate $\epsilon$.
In practice, we first calculate $\rave$ and $\robs$ from the data,
and then calculate $\rcal$ using Eq.(\ref{eq:r2ofr}). The bias is
estimated through
\be
b_r=\dfrac{\robs -\rcal}{2\langle r_{\rm obs}\rangle- {\pi \over 4}
\rund{{\rm e}^{\pi \langle r_{\rm obs}\rangle /2} -
{\rm e}^{-\pi \langle r_{\rm obs}\rangle /2}}}.
\elabel{ratiobias}
\ee
If a negative $b_r$ is obtained, we use $b_r = 2|\robs-\rcal|$ as an
empirical approximation.
\section{Numerical test with a non-singular isothermal ellipsoid model}

In this section we describe some simulations that we have
performed in order to test the behaviours of the estimators given in the
previous section. We model the halo surface mass density by a
non-singular isothermal ellipsoid (NIE) profile:
\be
\kappa(\theta)={\theta_{\rm E} \over 2\sqrt{\theta_c^2+\theta^2}},
\elabel{niehalo}
\ee
where the Einstein radius is $\theta_{\rm E}=6$ arcsec and the core
radius is taken to be $\theta_c=2$ arcsec. The redshifts of mock
sources are infinite. We have 50 flexion data points in each
realisation, which are randomly distributed on $1\times 1$ arcmin$^2$
``source plane''. Data inside 6 arcsec and outside 30 arcsec are
discarded (see Sect.3.1).

\subsection{Test of the weight function}

We generate 20 sets of mock data with different ellipticities
$\epsilon=0.03i$, $i=1,2,...20$. For each ellipticity, we use 50
realisations. For each realisation, we calculate the expected flexion
ratio in two ways. First, we calculate it directly using the mean of
the data; second, we calculate $\langle r\rangle$ through our weight
function (Eq.\ref{eq:weightr}). Then we estimate $\epsilon$ using
Eq.(\ref{eq:estimator1}). In Fig.\ts\ref{fig:weight}, we show the mean
estimate $\hat\epsilon$ over 50 realisations vs. the input values. The
solid line shows the equality line, the pluses and circles are
estimates with and without the weight function respectively.  The
error bars show the standard deviations. Both estimates closely trace
the input values, but the ones with weight function have smaller
variances, especially for large ellipticities.  The bottom panel shows
the relative error $|\delta\epsilon/\epsilon|$. One can see that the
biases with the weight function are significantly smaller, and all are
below $2\%$.  Hence throughout this paper the weight function has been
employed.

\subsection{The bias reduction}

We apply a simple model to generate noise in the data,
${\cal F}^{\rm obs}={\cal F}_1+n_{f1} + {\rm i}({\cal F}_2+n_{f2})$
(as in ES11). The intrinsic flexion noises $n_{f1},n_{f2}$ are
drawn from a Gaussian distribution independently, each component is
characterized by a standard deviation
$\sigma_{{\cal F}1}=\sigma_{{\cal F}2}=0.03$ arcsec$^{-1}$.

We first compare $\langle r^2\rangle$ using two methods. In
Fig.\ts\ref{fig:r2dif}, we show $\robs$ (pluses) and $\rcal$ (circles) of
different input $\epsilon$. For each $\epsilon$, we plot 5
realisations. The solid line is the theoretical prediction. Although
the noise is generated independent of $\epsilon$, the error and
difference between $\robs$ and $\rcal$ do increase with $\epsilon$.

\begin{figure}
  \centering
  \includegraphics[angle=0,scale=1.5]{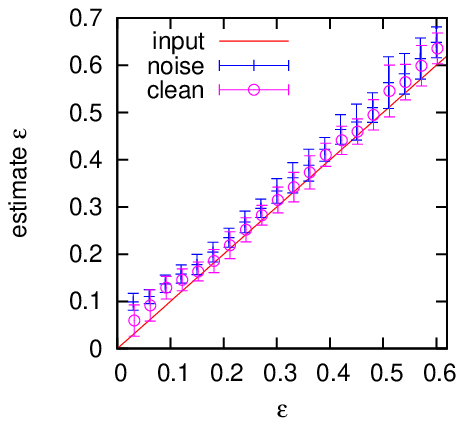}
  \includegraphics[angle=0,scale=1.5]{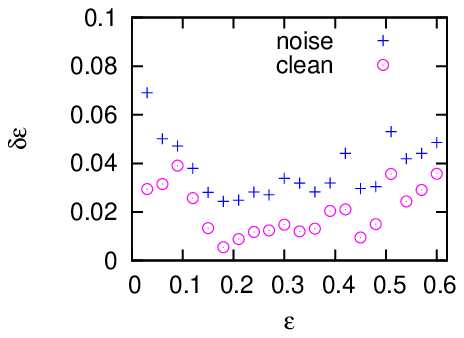}
  \caption{The top panel shows the comparison of the halo ellipticity
    estimator with the input ellipticity using the NIE model. The
    solid line shows the equality line.  The bottom panel shows the
    estimate errors. In both panels, the pluses are the average of
    estimate for 50 realisations with intrinsic noise while the
    circles are calculated with our clean method. }
  \label{fig:nierr2}
\end{figure}

Next we perform a test on our approach to bias reduction. 50 noise
realisations for each $\epsilon$ are generated. In each realisation,
$\rave$ and $\robs$ are calculated directly from the mock data, and
$\rcal$ are calculated using Eq. (\ref{eq:r2ofr}). In
Fig.\ts\ref{fig:nierr2} we compare the estimated values with the input
ones (solid line). The pluses are the average of 50 realisations using
Eq.(\ref{eq:estimator1}). We also estimate the bias $b_r$ using
Eq.(\ref{eq:delta}), and obtain an estimate from $\langle r^2_{\rm
  clean} \rangle$ using Eq.(\ref{eq:estimator2}). We need to point out
that for small $\epsilon$ (in case of our noise model, $\epsilon$
  $<0.1$), we adopt a modified approach since the linear approximation
(Eq.\ref{eq:delta}) is no longer valid. The results are shown by the
circles in Fig.\ts\ref{fig:nierr2}. Overall, the bias is reduced, but
for small $\epsilon$ we still have larger errors. The bias reduction
method is employed if not mentioned otherwise in Sect.3.

%
%%%%%%%%%%%%%%%%%%%%%%%%%%%
\bibliographystyle{mn2e}
\bibliography{../../../bib/refbooks,../../../bib/lens,../../../bib/flexion,../../../bib/refcos,../../../bib/shape}

\end{document}